\documentclass[english,aps, superscriptaddress, floatfix, 12pt]{revtex4}
\usepackage[T1]{fontenc}
\usepackage[latin1]{inputenc}
\usepackage{graphicx}
\usepackage{amssymb}
\usepackage{epsfig}
\usepackage{amsmath}
\usepackage{psfrag}

\makeatletter

\usepackage{babel}
\makeatother
\begin{document}

\title{Jet quenching parameter $\hat q$ in the stochastic QCD vacuum
with Landau damping}

\author{Dmitri Antonov, Hans-J\"urgen Pirner\\
{\it Institut f\"ur Theoretische Physik, Universit\"at Heidelberg,\\
Philosophenweg 19, D-69120 Heidelberg, Germany}}

\begin{abstract}
We argue that the radiative energy loss of a parton traversing the
quark-gluon plasma is determined by Landau damping of soft modes
in the plasma. Using this idea, we calculate the jet quenching parameter of a
gluon. The calculation is done in SU(3) quenched QCD within the
stochastic vacuum model. At the LHC-relevant temperatures, the result
depends on the gluon condensate, the vacuum correlation length, and
the gluon Debye mass. Numerically, when the temperature varies from
$T=T_c$ to $T=900{\,}{\rm MeV}$, the jet quenching parameter rises from $\hat q=0$
to approximately $1.8{\,}{\rm GeV^2}/{\rm fm}$.  We compare our results with the
predictions of perturbative QCD and other calculations.
\end{abstract}

\maketitle

\section{Introduction}

At RHIC and LHC energies, radiative energy loss is an important 
mechanism for jet energy loss in the quark-gluon
plasma~\cite{baier}. It is related to the mean transverse momentum $\left<p_\perp^2\right>$
which the parton acquires traversing the plasma. The momentum broadening of the parton is
proportional to the distance it travels, therefore the integrated 
total energy loss from radiation due to the acceleration is proportional to  
the distance travelled squared $L_\parallel^2$~\cite{baier}. The energy loss is parametrized by the 
jet quenching parameter $\hat q$ 
$$\Delta E=\frac{\alpha_s}{8}C_R\hat q L_\parallel^2,$$
where $C_R$ is the quadratic Casimir operator of the representation $R$ of the parton,
and $\alpha_s$ is the strong coupling constant at an appropriate scale.
The jet quenching parameter can be estimated in a dilute plasma perturbatively:
$$\hat q=\left<\sigma p_\perp^2\right>\rho\propto T^3 \int \frac{d\sigma}{d p_\perp^2}p_\perp^2 d^2p_\perp
\sim\alpha_s^2N_c^2T^3\ln\frac{1}{g}.$$
The perturbative temperature dependence of $\hat q$ is determined 
by the density of scattering partners $\rho \propto T^3$ and the differential transport Coulomb cross section,
which is cut off in the infra-red at the Debye mass.

A standard nonperturbative calculation of the 
jet quenching parameter 
can be done with the help of the dipole formalism~\cite{Kopeliovich_and_Huefner}.
In this formalism, a fake dipole of size $L_{\bot}$ is constructed from the 
partons in the $T$-amplitude and in the $T^{*}$-amplitude, the trajectories of which 
are displaced from each other by the distance $L_{\bot}$. The transport parameter $\left<\sigma p_\perp^2\right>$
can then be calculated from the cross section of this fake dipole. 
In the dipole model, for a dipole of size $L_{\bot}$ one has:
$$
\sigma(L_{\bot})={\rm tr}
\int d^2b \left<\left|V({\bf b})-V({\bf b}+{\bf L}_{\bot})\right|^2\right>=$$
$$={\rm tr} \int d^2b \left<2-V({\bf b})V^\dagger({\bf b}+{\bf L}_{\bot})-V^\dagger({\bf b})V({\bf b}+{\bf L}_{\bot})\right>.$$
Here
$$V({\bf b})={\cal P}{\,}\exp\left[igv_\mu\int_{-\infty}^{+\infty}d\tau A_\mu(x(\tau))\right]$$
is the Wilson line of a parton with the impact parameter ${\bf b}$ propagating with the 4-velocity $v_\mu=(1,{\bf v})$,
${\bf v}^2=1$, ${\bf v}{\bf b}=0$, and $x(\tau)=(\tau,{\bf b}+{\bf v}\tau)$~\footnote{Throughout the paper, we use the notation $A_\mu\equiv A_\mu^at^a$, where 
$t^a$'s are the generators of the group SU($N_c$) in a given representation $R$. The trace "tr" is normalized by dividing over the
trace of the unit matrix $\hat 1_R$ in the representation $R$. Therefore, with this definition, ${\rm tr}{\,}\hat
1_R=1$.}. The nonperturbative differential cross section to produce a parton with the transverse momentum $p_\perp$ is then
given by Fourier transform 
$$\frac{d\sigma}{dp_\perp^2}=\int d^2L_\perp
{\rm e}^{i{\bf p}_\perp{\bf L}_\perp}{\,}{\rm tr}\int d^2b 
\left<2-V({\bf b})V^\dagger({\bf b}+{\bf L}_{\bot})-V^\dagger({\bf b})V({\bf b}+{\bf L}_{\bot})\right>.$$
The expectation values $\left<\ldots\right>$ of the Wilson lines have to be evaluated with the target ground states. 
The excited states are summed over. The transport parameter can then be obtained by differentiation
$$
\left<\sigma p_\perp^2\right>=$$ 
$$=\int\frac{d^2p_\perp}{(2\pi)^2} \int d^2L_{\bot} \left(-\nabla_{\bot}^2
{\rm e}^{i{\bf p}_\perp{\bf L}_{\bot}}\right){\,}{\rm tr}\int d^2b
\left<2-V({\bf b})V^\dagger({\bf b}+{\bf L}_{\bot})-V^\dagger({\bf b})V({\bf b}+{\bf L}_{\bot})
\right>.$$
After a partial integration, one sees that 
because of the $p_\perp$-integration only the $L_\perp^2$-dependent part, i.e. the first-order term 
in the dipole cross section is relevant for $p_\perp$-broadening calculation.
When one takes the expectation value in the medium, the density of scattering partners in 
the medium enters. One can replace the traced products of Wilson lines, ${\rm tr}{\,}VV^\dagger$ and 
${\rm tr}{\,}V^\dagger V$, by two gauge-invariant Wilson loops, which facilitates the calculation. For example, one has
$${\rm tr}{\,}V({\bf b})V^\dagger({\bf b}+{\bf L}_\perp)={\rm tr}{\,}{\cal P}\exp\left\{igv_\mu\int_{-\infty}^{+\infty}
d\tau\left[A_\mu(x(\tau))-A_\mu(y(\tau))\right]\right\},$$
where $y(\tau)=(\tau,{\bf b}+{\bf L}_\perp-{\bf v}\tau)$. One can approximate two
long parallel lines $x(\tau)$ and $y(\tau)$ by a contour, which closes asymptotically at $\tau=\pm\infty$, where the interactions vanish. This yields the Wilson loop
$${\rm tr}{\,}V({\bf b})V^\dagger({\bf b}+{\bf L}_\perp)\simeq {\rm tr}{\,}{\cal P}\exp\left[ig\int_0^1ds\dot z_\mu(s)
A_\mu(z(s))\right]$$
defined at the united contour
$$z_\mu(s)=\left\{\begin{array}{rcl}
x_\mu(\tau(s)),~~ s\in\left(0,\frac12\right)\\
y_\mu(\tau(s)),~~ s\in\left(\frac12,1\right)
                  \end{array},
\right.~~ {\rm where}~~ \tau(s)=\tan(2\pi(s-1/4)).$$

At high energies, the $S$-matrix for dipole scattering is mostly real-valued 
because of purely absorptive interactions (Pomeron exchange). Other contributions have been
discussed in the literature~\cite{Nachtmann, book1}. In the calculation, which we will present in this paper, 
the $S$-matrix has both real- and imaginary-valued parts, therefore we have to guarantee the reality of the cross
section by using the real-valued part of the expectation value of the Wilson loop.
The real-valued part of the expectation value of the Wilson loop in the medium can be related
to the transport parameter $\hat q$ as follows
(see e.g. Ref.~\cite{Liu:2006he}):
\begin{equation}
\label{Wadj}
\left<{\rm Re}{\,}W^{\rm Mink}_{L_{\parallel}\times
L_{\perp}}\right>=\exp\left(-\frac{\hat q}{4\sqrt{2}}L_\parallel
L_\perp^2\right).
\end{equation}

As a tool to obtain the Wilson-loop expectation value we will use the so-called stochastic
vacuum model (SVM)~\cite{svm} in
Minkowski space-time. This model has been applied to high energy
hadron-proton and $\gamma^{*}$-proton scattering~\cite{ssp} 
with two Wilson loops representing projectile and target particles. 
Recently, a single Wilson loop near the light cone has 
also been studied within the SVM in
Ref.~\cite{pn}. This calculation is tricky since the 
expectation value of the Wilson loop on the light cone is strictly zero in vacuum,
the limiting procedure of approaching the light cone, however, can give the 
quark-antiquark confining interaction in the light-cone Hamiltonian.

\begin{figure}
\epsfig{file=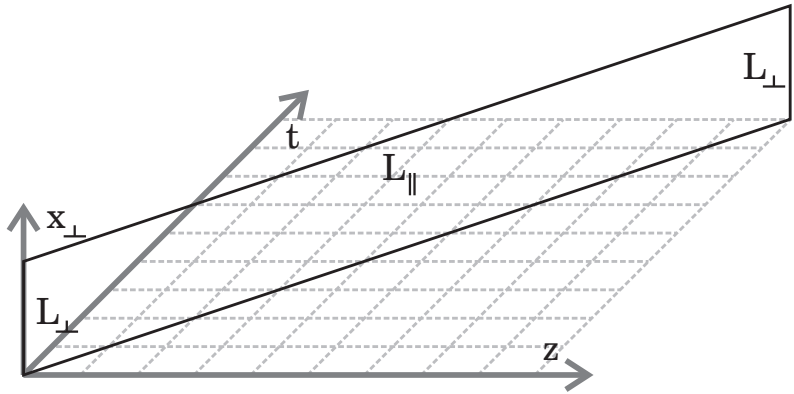, width=105mm}
\caption{\label{one} The contour of the Wilson loop ($L_\parallel\times L_\perp$) of a gluon.}
\end{figure}

In Fig.~\ref{one} we show the arrangement which we use for the following calculation. 
The long side of the contour of the Wilson loop is directed along the
light cone. The mean $\left<p_\perp^2\right>$ is
related to the transverse size $L_\perp$ of the contour.  A
counterintuitive property of Eq.~(\ref{Wadj}) is its exponential
fall-off, since normally an expectation value of a Wilson loop in
Minkowski space-time contains the imaginary unity $i$ in the argument
of the exponent~\cite{pn}. We will demonstrate that the exponential
fall-off is a consequence of Landau damping of soft modes in the
quark-gluon plasma and calculate through this mechanism the jet
quenching parameter $\hat q$.

Recently, a calculation of $\hat q$ at strong coupling has been done for
${\cal N}=4$ SYM in Refs.~\cite{Liu:2006he}, with the help of the
AdS/CFT-correspondence. It yields
$\hat q\propto T^3$, which is a consequence of conformal invariance. 
However, it may be that at LHC-relevant temperatures, the QCD conformal anomaly still
plays an important role. 
Lattice results on $(\varepsilon-3p) \ne 0$ indicate that 
the energy density  $\varepsilon$  and the pressure $p$  
deviate from pure Stefan-Boltzmann behavior.
For this reason, one may expect a violation of conformal invariance and an explicit dependence
of $\hat q$ on the gluon condensate, which arises in the SVM.
Of course, the chromo-electric gluon condensate vanished at high temperatures after
deconfinement, but there can remain a chromo-magnetic condensate, and nonvanishing
correlations between chromo-magnetic field strengths in the quark-gluon plasma. One of our main objectives in this paper is
to investigate their effect on the jet quenching parameter $\hat q$.

The paper is organized as follows. In section II, we review some basics  
about the SVM at $T=0$ and $T>T_c$. We also introduce there an effective
local formulation of the SVM, which will be technically important for
the subsequent analysis. In section III, we incorporate Landau damping
into the model and calculate the jet quenching parameter. In section IV
we summarize the main results of the paper.

\section{Generalities on Wilson loops and the SVM}

We start considering the expectation value of a Wilson loop in
Euclidean space-time at zero temperature ($T=0$).  In an arbitrary
representation $R$ of SU($N_c$), it reads $\Bigl<W(C)\Bigr>\equiv
\Bigl<{\rm tr}{\,}{\cal P}\exp\Bigl(ig\oint\limits_{C}^{} dx_\mu
A_\mu\Bigr)\Bigr>$. In the SVM, one uses the non-Abelian Stokes' theorem and
the cumulant expansion to write the Wilson-loop average as
follows~\cite{svm}:
\begin{equation}
\label{Wloop}
\Bigl<W(C)\Bigr>\simeq
{\rm tr}{\,}\exp\Biggl[-\frac{1}{2!}\frac{g^2}{4}\int\limits_{\Sigma(C)}^{}d\sigma_{\mu\nu}(x)
\int\limits_{\Sigma(C)}^{}d\sigma_{\lambda\rho}(x')\left<F_{\mu\nu}(x)F_{\lambda\rho}(x')\right>
\Biggr].
\end{equation}
Here, $\Sigma(C)$ is the surface encircled by the flat contour $C$. Averages involving gluonic field strengths 
are calculated in the SVM by using the two-point correlation 
function $\left<F_{\mu\nu}(x)F_{\lambda\rho}(x')\right>$, which
has been fitted to lattice data~\cite{Meggiolaro}. To simplify notations, we suppress
two phase factors between the points $x$ and $x'$ in the fundamental representation. Fixing the Fock-Schwinger gauge with the reference point in either of these two points, one can see that these phase factors are anyhow equal to $\hat 1_R$ when 
the path between the points $x$ and $x'$ is the straight line.
The approximate
equality in Eq.~(\ref{Wloop}) is due to the use of the cumulant
expansion, which gives the correct area law for the heavy quark-antiquark Wilson loop.
It is supported by lattice
data~\cite{9}-\cite{bali} and states that the amplitude of the
two-point irreducible gauge-invariant correlation function (cumulant)
of $F_{\mu\nu}$'s dominates over the amplitudes of higher-order
cumulants, which are therefore neglected. The factor $1/2!$ in
Eq.~(\ref{Wloop}) is due to the cumulant expansion, and the factor
$1/4$ is due to the non-Abelian Stokes' theorem.

Within the SVM, one parametrizes the nonperturbative part of
the two-point cumulant as follows~\cite{svm, ssdp}:
$$\left<F_{\mu\nu}^a(x)F_{\lambda\rho}^b(x')\right>=\delta^{ab}\frac{\left<(F_{\mu\nu}^a)^2\right>}{12(N_c^2-1)}
\Biggl\{\kappa(\delta_{\mu\lambda}\delta_{\nu\rho}-\delta_{\mu\rho}\delta_{\nu\lambda})D(u^2)+$$
\begin{equation}
\label{cor}
+\frac{1-\kappa}{2}\left[\partial_\mu\left(u_\lambda\delta_{\nu\rho}-u_\rho\delta_{\nu\lambda}\right)+
\partial_\nu\left(u_\rho\delta_{\mu\lambda}-u_\lambda\delta_{\mu\rho}\right)\right]D_1(u^2)\Biggr\}.
\end{equation}
Here, $D$ and $D_1$ are dimensionless functions of the distance $u=x-x'$ normalized by the condition  
$D(0)=D_1(0)=1$.  The gluon condensate defines the amplitude of this correlation function.
The function $D$ is responsible for the
confining properties of the non-Abelian gauge theory.  The function
$D_1$ describes Abelian-like self-interactions of the Wilson loop,
which do not lead to an expectation value of the form of
Eq.~(\ref{Wadj}). High-energy scattering data yield $\kappa=0.74$ in
the vacuum~\cite{ssdp}. We will take $\kappa=1$, i.e. disregard the small contribution of the function $D_1$
altogether. Lattice simulations~\cite{9}-\cite{bali} yield an exponential fall-off of the
function $D$:
$$D(u^2)={\rm e}^{-\mu|u|}$$
with the inverse vacuum correlation length~\cite{pisa2} 
$$\mu=894{\,}{\rm MeV}.$$
One can rewrite the Wilson-loop average using the
the surface tensor 
$$
\Sigma_{\mu\nu}(x)=\int_{\Sigma(C)}^{}d\sigma_{\mu\nu}(w(\xi))\delta(x-w(\xi)).
$$
The two-dimensional vector $\xi=(\xi_1,\xi_2)$, where $\xi_{1,2}\in[0,1]$, parametrizes differential elements
$d\sigma_{\mu\nu}$ on the surface $\Sigma(C)$, and $w_\mu(\xi)$ points towards each differential
surface element.
Using the correlation function $D$ one obtains the characteristic form of the 
Wilson-loop average: 
\begin{equation}
\label{Eff}
\Bigl<W(C)\Bigr>\simeq\exp\left[-\frac{C_R}{48(N_c^2-1)}
\cdot g^2\left<(F_{\mu\nu}^a)^2\right>\int d^4x\int d^4y\Sigma_{\mu\nu}(x)
{\rm e}^{-\mu|x-y|}\Sigma_{\mu\nu}(y)\right].
\end{equation}
In particular, for a
contour $C$ whose temporal extension is much larger than its spatial
extension, one obtains an area law depending on the
string tension in the fundamental representation~\cite{ssdp, aes}
\begin{equation}
\label{Sigma}
\sigma=\frac{\pi C_F}{12(N_c^2-1)}\frac{g^2
\left<(F_{\mu \nu}^a)^2\right>}{\mu^2}~~ {\rm with}~~ C_F=\frac{N_c^2-1}{2N_c}.
\end{equation}
For $N_c=3$, we take the standard value of the string tension 
$\sigma=(440{\,}{\rm MeV})^2$ to estimate the gluon condensate
$g^2\left<(F_{\mu\nu}^a)^2\right>\simeq3.55{\,}{\rm GeV}^4$.

We see from Eq.~(\ref{Eff}) that the SVM essentially suggests a
representation of the Wilson-loop average in terms of an effective
local field theory of the field strength tensor $F_{\mu\nu}^a$:
\begin{equation}
\label{wef}
\Bigl<W(C)\Bigr>= {\rm tr}{\,}\int {\cal D}F_{\mu\nu}^a{\rm e}^{-S_{\rm Eucl}[F]}
\end{equation}
with the action
\begin{equation}
\label{Seff}
S_{\rm Eucl}[F]=\frac12\int d^4x\left[F_{\mu\nu}^a{\cal K}(x)F_{\mu\nu}^a+iF_{\mu\nu}^at^a\Sigma_{\mu\nu}\right].
\end{equation}
Here, ${\cal K}$ is a well-defined local operator, which 
reproduces Eq.~(\ref{Eff}) after the Gaussian integration in Eq.~(\ref{wef}):
\begin{equation}
\label{Eff1}
\Bigl<W(C)\Bigr>=\exp\left[-\frac{C_R}{8}\int d^4x\int
d^4y\Sigma_{\mu\nu}(x){\cal K}^{-1}(x-y)\Sigma_{\mu\nu}(y)\right],
\end{equation}
where ${\cal K}^{-1}(x)$ is a Green function of the operator ${\cal K}$: ${\cal K}{\cal K}^{-1}=\delta(x)$.
Comparing Eqs.~(\ref{Eff}) and (\ref{Eff1}), we find that the kernel ${\cal K}^{-1}$ has an exponential fall-off:
\begin{equation}
\label{kmin1}
{\cal K}^{-1}(x)=\frac{g^2\left<(F_{\mu\nu}^a)^2\right>}{6(N_c^2-1)}{\rm e}^{-\mu|x|}.
\end{equation}
The operator ${\cal K}$ itself can then be obtained through Fourier
transformation. In the coordinate representation, it reads
\begin{equation}
\label{calK}
{\cal K}(x)=\frac{N_c^2-1}{2\pi^2}\frac{\mu^4}{g^2\left<(F_{\mu\nu}^a)^2\right>}\left(1-\frac{\partial^2}{\mu^2}\right)^{5/2}.
\end{equation}

At finite temperature ($T\ne 0$), the $O(4)$ space-time symmetry is
broken down to the spatial $O(3)$ symmetry, and the
correlation function~(\ref{cor}) splits into the following three functions
$$\left<E_i^a(x)E_k^b(x')\right>,~~ \left<E_i^a(x)B_k^b(x')\right>,~~ \left<B_i^a(x)B_k^b(x')\right>,$$
which were
simulated on the lattice in Refs.~\cite{pisa1, pisa2}. In the deconfined phase ($T>T_c$), the chromo-electric 
condensate vanishes, and so does the correlation function $\left<E_i^a(x)E_k^b(x')\right>$.
Furthermore, the amplitude of the mixed electric-magnetic correlation function is
by an order of magnitude smaller than the amplitude of the magnetic-magnetic correlation function~\cite{pisa1}.
For this reason, we disregard the correlation function $\left<E_i^a(x)B_k^b(x')\right>$ with respect to the function  $\left<B_i^a(x)B_k^b(x')\right>$, which can be parametrized as follows~\footnote{The normalization of the function $D^B$ can be proven by taking the $(T\to 0)$-limit and imposing for the correlation function 
$\left<E_i^a(x)E_k^b(x')\right>$ the same parametrization~(\ref{BB}) with some other function $D^E$ instead of 
$D^B$. Due to the unbroken $O(4)$-invariance at $T=0$, one has $D^E(0)=D^B(0)$. On the other hand, using the parametrization~(\ref{BB}), one has 
$$\left<(F_{\mu\nu}^a)^2\right>=2\sum\limits_{i=1}^{3}\left[\left<(E_i^a)^2\right>+\left<(B_i^a)^2\right>\right]=
\frac12\left<(F_{\mu\nu}^a)^2\right>\left[D^E(0)+D^B(0)\right].$$
This proves that $D^B(0)=1$.}:
\begin{equation}
\label{BB}
\left<B_i^a(x)B_k^b(x')\right>=\frac{\left<(F_{\mu\nu}^a)^2\right>_T}{12(N_c^2-1)} 
\delta^{ab}\delta_{ik}D^B(u^2),
\end{equation} 
\begin{equation}
\label{BB1}
D^B(u^2)={\rm e}^{-\mu(T)|u|}.
\end{equation}
The temperature dependencies of the gluon condensate and of the inverse vacuum correlation length will be discussed in the 
next section.

\section{Evaluation of $\hat q$ through Landau damping}

We want to calculate now the Wilson-loop average in the gluon plasma,
i.e. in a thermal environment.  At finite temperature, in the
Euclidean space-time, the contour of the loop has the orientation
shown in Fig.~\ref{one}, namely it is oriented at $45^\circ$ in the
$(x_4,x_3)$-plane.  Due to the $x_4$-periodicity, the contour
$C=L_\parallel\times L_\perp$ effectively splits into pieces, whose
extensions along the 3rd and the 4th axes are $\beta\equiv 1/T$. Such
pieces will be referred to as strips.  A point lying on the surface of
the strip closest to the origin can be parametrized by a
vector-function
\begin{equation}
\label{Ww}
w_\mu(\xi_1,\xi_2)=\beta\xi_1t_\mu+L_\perp\xi_2r_\mu,
\end{equation}
$$
t_\mu=(0,0,1,1),~ r_\mu=(1,0,0,0),~ \xi_{1,2}\in[0,1].$$
Since we have argued to disregard the mixed correlator, only the
contribution of the chromo-magnetic field to the effective action
should be taken into account, which yields [cf. Eq.~(\ref{Seff})]
\begin{equation}
\label{seucl}
S_{\rm Eucl}[B]=\int d^4x\left(B_2^a{\cal K}B_2^a+iB_2^at^a\Sigma_{13}\right).
\end{equation}
The Wilson-loop average of one strip reads
\begin{equation}
\label{one1}
\left<W_{\rm 1-strip}^{\rm Eucl}\right>=
{\rm tr}{\,}\int {\cal D}B_2^a{\rm e}^{-S_{\rm Eucl}[B]}=\exp\left[-\frac{C_R}{4}\int d^4x\int d^4y\Sigma_{13}(x)
{\cal K}^{-1}(x-y)\Sigma_{13}(y)\right].
\end{equation}
Note that, for ${\cal K}^{-1}$ given by Eq.~(\ref{kmin1}), this result
indeed agrees with the one following from a direct application of the
cumulant expansion,
$$\left<W_{\rm 1-strip}^{\rm Eucl}\right>={\rm
tr}{\,}\left<\exp\left(i\int
d\sigma_{13}B_2^at^a\right)\right>\simeq$$
$$\simeq{\rm tr}{\,}\exp\left[-\frac{g^2}{2}\int d\sigma_{13}(w)\int
d\sigma_{13}(w')t^at^b\left<B_2^a(w)B_2^b(w')\right>\right],$$ when
one uses $\left<B_2^a(w)B_2^b(w')\right>$ in the form of
Eq.~(\ref{BB}).

In the full Wilson-loop average $\left<W_{L_{\parallel}\times L_{\perp}}^{\rm Eucl}\right>$, 
also interactions between different strips take place besides self-interactions of one strip. 
Let $\chi_k$ denote the interaction between two strips separated from each other by the distance $\beta k$:
$$\chi_k=\frac{C_R}{4}\int d\sigma_{13}(w)\int d\sigma_{13}(w'){\cal K}^{-1}(w-w'),$$
where $w_\mu'=w_\mu(\xi_1',\xi_2')+(0,0,\beta k,0)_\mu$.
Here, the relative distance between the points, $|\Delta w(k)|\equiv|w-w'|$, is 
$$
|\Delta w(k)|=$$
\begin{equation}
\label{Dw}
=\left\{\left[\beta(\xi_1'-\xi_1)+k\beta\right]^2+\left[\beta(\xi_1'-\xi_1)\right]^2+{\cal O}
\left[L_\perp^2(\xi_2'-\xi_2)^2\right]\right\}^{1/2}\simeq
\beta\sqrt{k^2+2kx+2x^2},
\end{equation}
$$x\equiv\xi_1'-\xi_1,~~ x\in[-1,1].$$ 
In the final form of
Eq.~(\ref{Dw}), we disregard ${\cal
O}\left[L_\perp^2(\xi_2'-\xi_2)^2\right]$, since $L_\perp\ll\beta$. If
this term had been retained, the final result for $\hat q$ would
depend on the size of the color dipole, $L_\perp$.  The full number of
strips (i.e. the maximal value of $k$) is
\begin{equation}
\label{Nn}
n\equiv\frac{L_\parallel}{\beta\sqrt{2}}.
\end{equation} 
Therefore, the overall contribution to the Wilson-loop average reads
\begin{equation}
\label{sums}
-\ln\left<W_{L_\parallel\times L_\perp}^{\rm Eucl}\right>=
\sum_{i=0}^{n-1}\sum_{k=0}^{i}\chi_k=\sum\limits_{k=0}^{n-1}(n-k)\chi_k.
\end{equation}

\begin{figure}
\epsfig{file=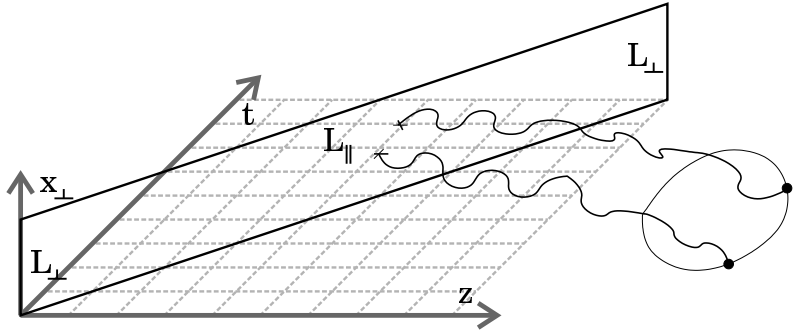, width=105mm}
\caption{\label{two} The Wilson-loop surface interacting with the hard gluons (drawn curve) 
through the soft ones (wavy curves). The imaginary part of the soft-gluon polarization operator corresponds to
Landau damping. For simplicity, the splitting of the surface into strips is not depicted.}
\end{figure}

For the transport parameter $\hat q$ we need a calculation of the real-valued part of the adjoint 
Wilson-loop expectation value in Minkowski space-time. We would like to discuss this expectation value first, 
before we make the analytic continuation. In Minkowski space-time, in general 
$\left<W^{\rm Mink}(C)\right>=\eta{\rm e}^{2i\delta}$. If one interpretes the Wilson loop as an interaction 
of the fake dipole with the medium, then there can be two medium effects. The fake-meson interaction can generate 
real- and imaginary-valued parts of the $S$-matrix. Absorptive effects
[cf. Eq.~(\ref{Wadj})], namely the reduction of the intensity of the beam when it traverses the medium, are described 
by $\eta$. In addition, there is  
real-valued phase shift $\delta$ of the wave-function of the projectile, which it 
acquires propagating through the target. In our calculation, both terms will be present, but only $\eta$ 
will contribute to $\left<\sigma p_\perp^2\right>$. That is because the real-valued part of the 
$S$-matrix $\propto\cos(2\delta)$ differentiated twice with respect to the dipole size $L_\perp$ and evaluated then at 
$L_\perp=0$ vanishes. With these preliminaries in mind we can now make the 
important analytic continuation to Minkowski space-time.
In Minkowski space-time, the action~(\ref{seucl}) goes over to 
\begin{equation}
\label{smink}
S_{\rm Mink}[B]=i\int d^4x\left(B_2^a{\cal K}B_2^a+B_2^at^a\Sigma_{13}\right),
\end{equation}
and Eq.~(\ref{one1}) becomes
$$
\left<{\rm Re}{\,}W_{\rm 1-strip}^{\rm Mink}\right>=
{\rm tr}{\,}{\rm Re}{\,}\int {\cal D}B_2^a{\rm e}^{-S_{\rm Mink}[B]}=$$
\begin{equation}
\label{15}
={\rm Re}{\,}\exp\left[i\frac{C_R}{4}\int d^4x\int d^4y\Sigma_{13}(x)
{\cal K}^{-1}(x-y)\Sigma_{13}(y)\right].
\end{equation}
Until now, this average does not have an exponential fall-off. The exponential fall-off only appears when we 
take into account the scattering partners in the medium as external sources. The correlation functions must mediate 
an interaction with the gluons in the heat bath, otherwise there is no absorptive scattering. These gluons 
can be summed over, see Fig.~\ref{two}. They will then appear as a gluon polarization insertion into the 
correlation function of the field strengths. The imaginary-valued part of this polarization operator comes from the 
on-shell gluons in the intermediate state, which contribute to the absorptive phase. Since the imaginary-valued part
of the gluon self-energy is related to the soft modes of the quark-gluon plasma~\cite{land}, one sees that the 
energy loss is essentially linked to Landau damping of gluons. The soft background gluons with energies 
$\omega\equiv |p_0|\ll|{\bf p}|$, parametrized by the SVM correlation function, can be absorbed 
by thermal gluons, i.e. gluons from the thermal bath. Accordingly, the stochastic QCD vacuum becomes modified 
in a way, which is not accessible in Euclidean simulations on the lattice and, therefore, is not encoded 
in the parametrization of the $D^B$-function, Eq.~(\ref{BB1}).
The imaginary part of the thermal loop for the gluon
self-energy reads in the hard-loop approximation~\cite{land, polariz}
$${\rm Im}~ \Pi_{ij}(\omega,{\bf p})=-\pi m_D^2\omega\int\frac{d\Omega}{4\pi}n_i n_j\delta(\omega-{\bf n}{\bf p})=$$
\begin{equation}
\label{impart}
=-\frac{\pi m_D^2\omega}{2|{\bf p}|}\left[\frac{\omega^2}{{\bf
p}^2}\cdot \frac{p_ip_j}{{\bf
p}^2}+\frac12\left(1-\frac{\omega^2}{{\bf
p}^2}\right)\cdot\left(\delta_{ij}- \frac{p_ip_j}{{\bf
p}^2}\right)\right].
\end{equation}
Here $m_D=gT\sqrt{N_c/3}$ is the Debye mass of
the hard thermal gluon in quenched QCD under study. The separation of the 
momentum scales of the correlation function and the thermal gluons is rather subtle. The correlation function 
contains spatial momenta of order $\mu$, but very small energies $\omega\ll m_D$, whereas the thermal heat bath
contains gluons with energies and spatial momenta of order $T$.
For ultrasoft space-like gluons with
$\omega^2\ll{\bf p}^2$, only the transverse part of Eq.~(\ref{impart}) survives:
$$
{\rm Im}~ \Pi_{ij}(\omega,{\bf p})\to\left(p_ip_j-{\bf p}^2\delta_{ij}\right)\cdot{\cal P}(\omega,{\bf p}),$$
\begin{equation}
\label{imP}
{\cal P}(\omega,{\bf p})=\frac{\pi m_D^2\omega}{4|{\bf p}|^3}.
\end{equation}
The linear component of the field-strength tensor
($\sim\partial_i A_j^a$) satisfies the relation
$$F_{ij}^{a{\,}({\rm lin})}({\bf p})F_{ij}^{b{\,}({\rm lin})}(-{\bf
p})=
2\left({\bf p}^2\delta_{ij}-p_ip_j\right)A_i^a({\bf p})A_j^b(-{\bf p}).$$
Polyakov~\cite{book} has argued that, in the course of
integration over hard modes, the cubic ($\sim A^3$) and the quartic
($\sim A^4$) terms appear in such a way that they supplement the
square of the field-strength tensor to the standard non-Abelian
form. For this reason, the spatial part of the effective
action~(\ref{Seff}) in Minkowski space-time may be modified by ${\rm
Im}~ \Pi_{ij}(\omega,{\bf p})$:
$$
S_{\rm Mink}[F]=\frac{i}{2}\int d^4x\left[F_{ij}^a{\cal K}(x)F_{ij}^a+F_{ij}^at^a\Sigma_{ij}\right]\to$$
\begin{equation}
\label{change}
\to\frac{i}{2}\int d^4x\left\{F_{ij}^a\left[{\cal K}(x)-i{\cal P}(x)\right]F_{ij}^a+F_{ij}^at^a\Sigma_{ij}\right\}.
\end{equation}
Accordingly, Eq.~(\ref{15}) changes as 
$$
\left<{\rm Re}{\,}W_{\rm 1-strip}^{\rm Mink}\right>={\rm Re}{\,}\exp\left[i\frac{C_R}{4}\int d^4x\int d^4y\Sigma_{13}(x)\left(
\frac{1}{{\cal K}-i{\cal P}}\right)_{xy}\Sigma_{13}(y)\right]=$$
\begin{equation}
\label{change1}
={\rm Re}{\,}\exp\left[-\frac{C_R}{4}\int d^4x\int d^4y\Sigma_{13}(x)\left(\frac{{\cal P}-i{\cal K}}{{\cal K}^2+
{\cal P}^2}\right)_{xy}\Sigma_{13}(y)\right].
\end{equation}
As was discussed, the term proportional to $i{\cal K}$ in Eq.~(\ref{change1}) describes the 
non-absorptive part of the interaction between the fake dipole and the medium. It does not contribute 
to $\hat q$ and will henceforth be disregarded. The remaining real-valued part of Eq.~(\ref{change1}) describes 
jet quenching. The interaction of the Wilson-loop surface with the hard gluons is depicted in Fig.~\ref{two}.
There, the closed line is the hard thermal loop, and the wavy lines are soft gluons, which mediate the 
interactions. The chromo-magnetic condensate and the hard gluons represent soft and hard components coexisting in the 
gluon plasma.

To handle the complicated Eqs.~(\ref{change}) and (\ref{change1}), we have
approximated ${\cal P}(\omega,{\bf p})$ by a function ${\cal P}(p)$ of
the full 4-momentum in the following way~\footnote{For brevity, we use
the same notations for a function or an operator and its Fourier
image. What is implied in each particular case is clearly defined by
the corresponding argument.}. We denote $\omega=\zeta|{\bf p}|$, where $\zeta\ll 1$. Then 
$|{\bf p}|=\sqrt{\frac{-p^2}{1-\zeta^2}}$, and we have 
$$
{\cal P}(p)=-\frac{M^2(T)}{p^2},$$
\begin{equation}
\label{zeta}
M^2(T)\equiv\frac{\pi\zeta(1-\zeta^2)}{4}m_D^2(T).
\end{equation}
An important property of the SVM for high-energy scattering in Minkowski space-time~\cite{ssp, ssdp} is the same exponential fall-off with transverse distance of the 
surface-surface correlations as in the Euclidean space-time [cf. Eq.~(\ref{15})]. For this reason, one may calculate the operator $\left(\frac{{\cal P}}{{\cal K}^2+{\cal P}^2}\right)_{xy}$ of interest in the Euclidean space-time, where 
${\cal P}(p)=\frac{M^2(T)}{p^2}$ and, according to Eq.~(\ref{calK}),
$${\cal K}^{-2}(p)=\frac{{\cal N}(T)}{[p^2+\mu^2(T)]^5}.$$ 
In this equation,
$${\cal N}(T)\equiv\left(\frac{2\pi^2\mu(T)g^2\left<(F_{\mu\nu}^a)^2\right>_T}{N_c^2-1}\right)^2$$
contains the temperature-dependent gluon condensate $g^2\left<(F_{\mu\nu}^a)^2\right>_T$ and the 
inverse vacuum correlation length $\mu(T)$. We therefore obtain
$$
\left<{\rm Re}{\,}W_{\rm 1-strip}^{\rm Mink}\right>=$$
\begin{equation}
\label{W01}
=\exp\left[-\frac{C_RM^2(T)}{4}\int d\sigma_{13}(w)\int d\sigma_{13}(w')
\int\frac{d^4p}{(2\pi)^4}{\rm e}^{ip(w-w')}\frac{p^2}{p^4{\cal K}^2(p)+M^4(T)}\right].
\end{equation}
One can now reduce the 4-dimensional momentum integration to a 1-dimensional integration by using the formula
$$\int\frac{d^4p}{(2\pi)^4}{\rm e}^{ip(w-w')}f(|p|)=\frac{1}{4\pi^2|w-w'|}\int_0^\infty dpp^2J_1(p|w-w'|)f(p)$$
with the $J_1$-Bessel function. According to Eq.~(\ref{Ww}), the product of two infinitesimal surface elements reads
$$d\sigma_{13}(w)d\sigma_{13}(w')=(\beta L_\perp)^2d^2\xi d^2\xi'.$$
Furthermore, one can simplify the area-area correlation inside one strip $(k=0)$, 
where the distance between the points depends only on the variable $x$ introduced in Eq.~(\ref{Dw}):
$$|w-w'|=|\Delta w(0)|=\beta|x|\sqrt{2}.$$
Equation~(\ref{W01}) yields the following expression:
\begin{equation}
\label{W1}
\chi_0=-\ln \left<{\rm Re}{\,}W_{\rm 1-strip}^{\rm Mink}\right>=\frac{C_R{\cal N}(T)M^2(T)}{16\pi^2\sqrt{2}}\beta L_\perp^2
\int_{-1}^{1}\frac{dx}{|x|}\int_0^\infty\frac{dpp^4J_1(p\beta|x|\sqrt{2})}{p^4(p^2+\mu^2(T))^5+{\cal N}(T)M^4(T)}.
\end{equation}
The $L_\perp$-dependence of this result is consistent with the color transparency
of a color-neutral dipole of size $L_\perp$ interacting with the gluon
plasma. Furthermore, the $x$-integral in Eq.~(\ref{W1}) does not diverge at $x=0$, since this singularity is cancelled by the 
Bessel function.

By using Eqs.~(\ref{Wadj}), (\ref{Nn}), and (\ref{sums}), we can now determine $\hat q$:
$$
\hat q=-\frac{4\sqrt{2}}{L_\parallel L_\perp^2}\ln\left<{\rm Re}{\,}
W^{\rm Mink}_{L_{\parallel}\times L_{\perp}}\right>= \frac{4}{\beta
L_\perp^2}\frac{1}{n}\sum\limits_{k=0}^{n-1}(n-k)\chi_k,
$$
where 
$$\chi_k=\frac{C_R{\cal N}(T)M^2(T)}{16\pi^2}\beta L_\perp^2
\int_{-1}^{1}\frac{dx}{\sqrt{k^2+2kx+2x^2}}\int_0^\infty\frac{dpp^4J_1(p\beta\sqrt{k^2+2kx+2x^2})}{p^4(p^2+\mu^2(T))^5+
{\cal N}(T)M^4(T)}.$$
Here, we have used Eq.~(\ref{Dw}) for the distance $|\Delta w(k)|$ between two points belonging to the 
strips separated from each other by the interval $\beta k$. Our final result reads

\begin{equation}
\label{final}
\hat q=\frac{C_R{\cal N}(T)M^2(T)}{4\pi^2n}\sum\limits_{k=0}^{n-1}(n-k)\int_{-1}^{1}\frac{dx}{\sqrt{k^2+2kx+2x^2}}
\int_0^\infty
\frac{dpp^4J_1\left(p\beta\sqrt{k^2+2kx+2x^2}\right)}{p^4(p^2+\mu^2(T))^5+{\cal N}(T)M^4(T)}.
\end{equation}
 

Let us consider a gluon traversing the medium, for which $R$ is the
adjoint representation, and $C_R=N_c=3$. Furthermore, we assume that the critical temperature $T_c=270{\,}{\rm MeV}$
in SU(3) quenched theory~\cite{tc}. The temperature behavior of $\hat q$ 
and the growth of $\hat q$ with $T$ depend on the functions 
$g^2\left<(F_{\mu\nu}^a)^2\right>_T$, $\mu(T)$, and $M(T)$. To see whether 
this dependence is strong or not, we determine $\hat q(T)$ for 
two possible choices (I and II) of these functions. 

In case I, the nonperturbative value of the thermal strong coupling $g(m_D,T)$ 
is fixed by self-consistency~\cite{pb}: $g(m_D,T)=2.5$.
This choice defines $M(T)$ through Eq.~(\ref{zeta}) with $m_D=g(m_D,T)T$. As suggested by the 
{\it low-temperature} lattice data~\cite{pisa1, pisa2},
the inverse vacuum correlation length can be taken constant up to the temperature of dimensional reduction~\cite{tdr}, 
$T_{\rm d.r.}\simeq2T_c=540{\,}{\rm MeV}$. Above this temperature, 
all dimensionful quantities become proportional to the corresponding power of $T$.
Therefore, $\mu(T)={\rm const}$ until $T_{\rm d.r.}$, whereas $\mu(T)\propto T$ at $T>T_{\rm d.r.}$. The
proportionality coefficient in the last equation can be fixed by the continuity of
$\mu(T)$ across $T_{\rm d.r.}$: 
$$\mu(T)=1.66T~~ {\rm at}~~ T>T_{\rm d.r.}.$$ 
The temperature dependence of the chromo-magnetic condensate was calculated in Ref.~\cite{nik}:
$$g^2\left<(F_{\mu\nu}^a)^2\right>_T=g^2\left<(F_{\mu\nu}^a)^2\right>
\coth\left(\frac{\mu}{2T}\right)~~ {\rm at}~~ T_c<T<T_{\rm d.r.}.
$$
It is also nearly constant up to $T_{\rm d.r.}$.
At $T>T_{\rm d.r.}$, $g^2\left<(F_{\mu\nu}^a)^2\right>_T\propto T^4$, and the proportionality coefficient
again follows from continuity of
$g^2\left<(F_{\mu\nu}^a)^2\right>_T$ across $T_{\rm d.r.}$: 
$$g^2\left<(F_{\mu\nu}^a)^2\right>_T=61.46T^4~~ {\rm at}~~ T>T_{\rm d.r.}.$$

\begin{figure}
\psfrag{GT}{$g^2\left<(F_{\mu\nu}^a)^2\right>_T{\,}[{\rm GeV}^4]$}
\epsfig{file=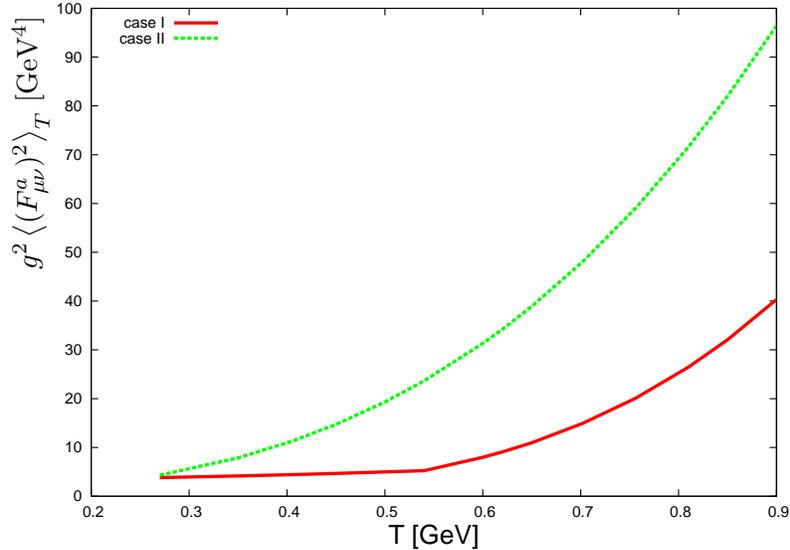, width=105mm}
\caption{\label{ra3} The chromo-magnetic condensate as a function of temperature at $T_c\le T\le 900{\,}{\rm MeV}$ 
in cases I and II.}
\end{figure}

In case II, we define the parameter $M(T)$ through the perturbative 1-loop strong coupling~\cite{tdr}: 
$$g^{-2}(T)=2b_0\ln\frac{T}{\Lambda},~~ {\rm where}~~  
b_0=\frac{11N_c}{48\pi^2}\Bigr|_{N_c=3}=\frac{11}{16\pi^2},~~ \Lambda\simeq0.104T_c.$$ 
Furthermore, {\it high-temperature} lattice data~\cite{tdr} suggest the parametrization 
$$\mu(T)=1.04 g^2(T)T,$$ 
where the coefficient has been fixed 
by the value $\mu(T_c)=\mu$. The same lattice data suggest also a temperature-dependent spatial string tension 
$$\sigma(T)=[cg^2(T)T]^2,~~ {\rm where}~~ c=0.566.$$ 
Note that such a value of the coefficient $c$ corresponds to a spatial string tension
$\sigma(T_c)=(485{\,}{\rm MeV})^2$,
that is slightly larger than the standard zero-temperature value $\sigma(0)=(440{\,}{\rm MeV})^2$.
The temperature-dependent chromo-magnetic condensate can then be obtained from Eq.~(\ref{Sigma}) at $N_c=3$: 
$g^2\left<(F_{\mu\nu}^a)^2\right>_T=\frac{72}{\pi}\mu^2(T)\sigma(T)$. For illustration, we present the chromo-magnetic 
condensates for both cases I and II in Fig.~\ref{ra3}.

\begin{figure}
\psfrag{qT}{$\hat q{\,}[{\rm GeV}^2/{\rm fm}]$}
\epsfig{file=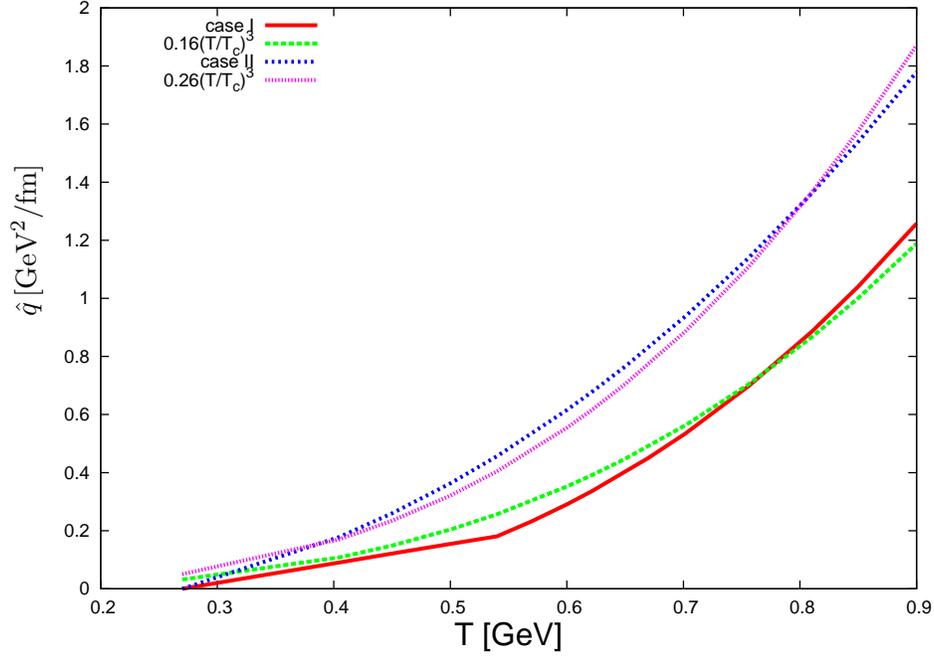, width=125mm}
\caption{\label{ra4} The jet quenching parameter $\hat q(T)$ in case I (lower full curve) and II (upper dashed curve), and the interpolating fitted curves $\sim T^3$ in both cases.}
\end{figure}

\begin{figure}
\psfrag{qT1}{$\hat q{\,}[{\rm GeV}^2/{\rm fm}],~~ {\rm case}~ {\rm I}$}
\epsfig{file=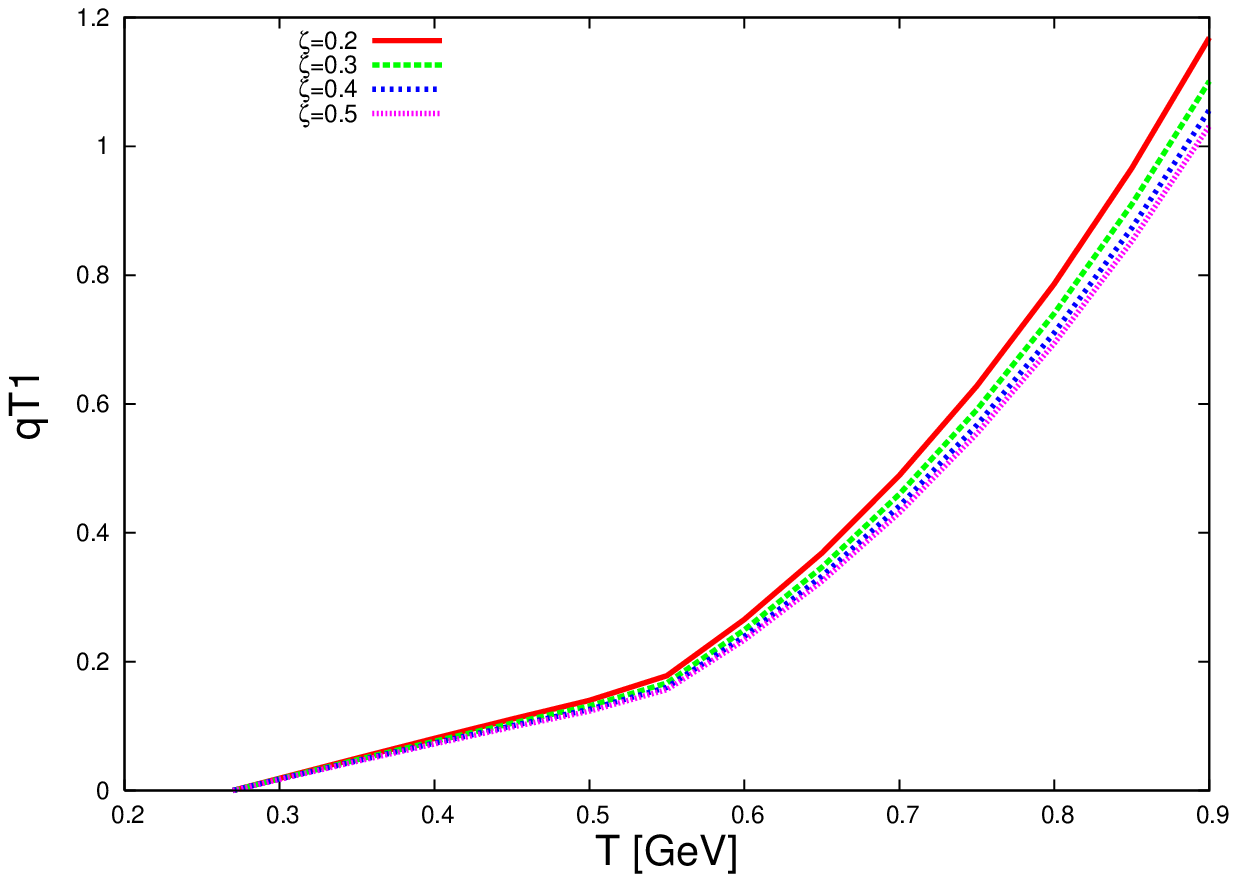, width=125mm}
\caption{\label{ra5} The jet quenching parameter $\hat q(T)$ in case I for various values of the parameter $\zeta$.}
\end{figure}

\begin{figure}
\psfrag{qT2}{$\hat q{\,}[{\rm GeV}^2/{\rm fm}],~~ {\rm case}~ {\rm II}$}
\epsfig{file=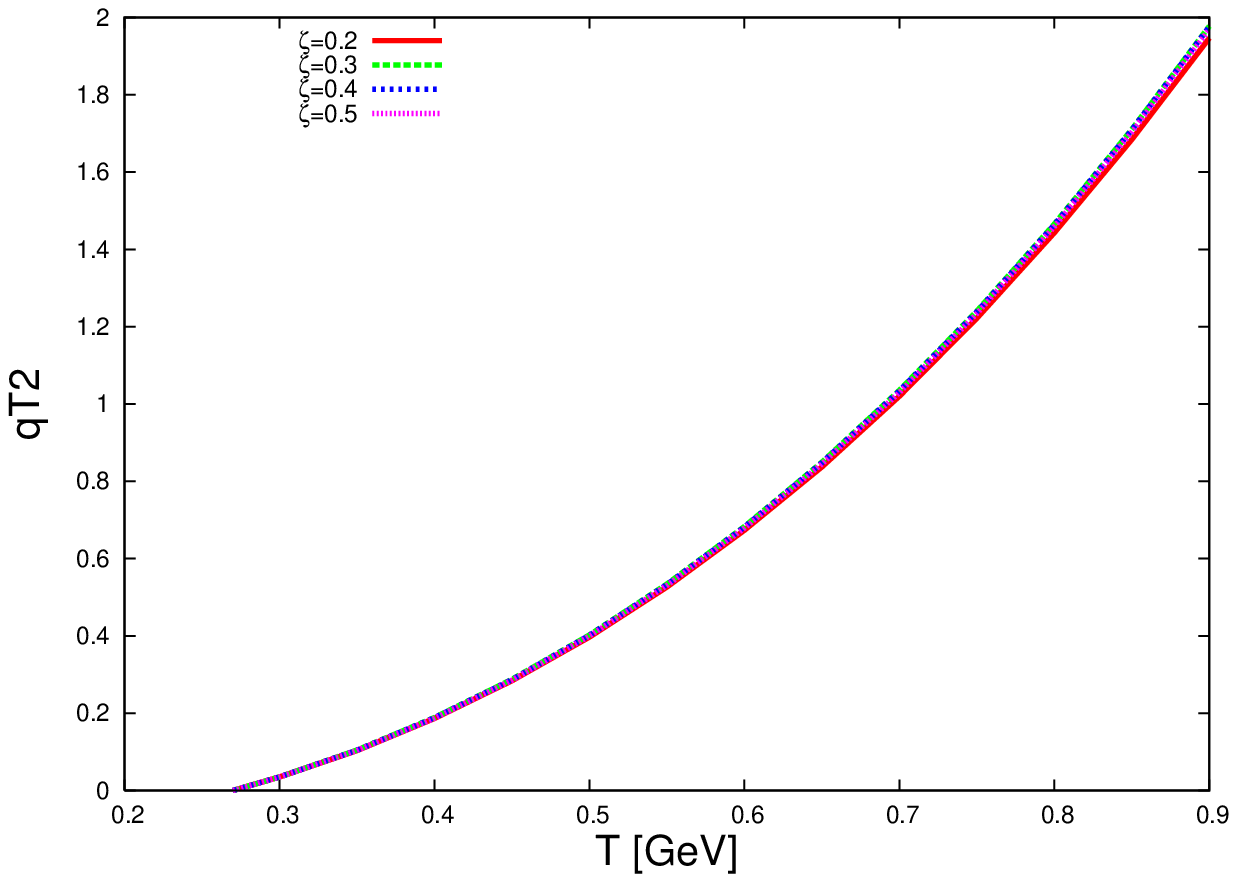, width=125mm}
\caption{\label{ra6} The jet quenching parameter $\hat q(T)$ in case II for various values of the parameter $\zeta$.}
\end{figure}

Because of the much smaller density of scattering partners in the hadronic phase, typical values of 
$\hat q$ at $T<T_c$ are $\sim 0.01{\,}{\rm GeV}^2/{\rm fm}$, which is by two orders of magnitude smaller than 
in the quark-gluon plasma~\cite{zz}. For this reason, we set $\hat q(T)=\hat q(T)-\hat q(T_c)$, i.e. 
$\hat q(T_c)=0$. Furthermore, we truncate $n=10\gg 1$ and use a small value $\zeta=0.1\ll 1$. We then obtain $\hat q$ as 
a function of temperature in the interval 
$270{\,}{\rm MeV}\le T\le 900{\,}{\rm MeV}$ for both cases I and II, cf. Fig.~\ref{ra4}. Both calculations
of $\hat q$ follow a temperature dependence $\propto T^3$. We find 
$$\hat q(T)_{\rm case{\,}I}^{\rm fit}=0.16(T/T_c)^3{\,}{\rm GeV^2/fm},~~ 
\hat q(T)_{\rm case{\,}II}^{\rm fit}=0.26(T/T_c)^3{\,}{\rm GeV^2/fm}.$$
These fits are inspired by Eq.~(\ref{final}). Indeed, counting the powers of $T$ in this equation at high temperatures, 
one sees that
$$\hat q\propto T^3~~ {\rm at}~~ T\gtrsim T_{\rm d.r.}.$$ 
Furthermore, at $T=900{\,}{\rm MeV}$ we have tested the sensitivity of our results against an increase of the 
number of strips $n$. An increase of the number of strips leads to a small increase of $\hat q(T)$, namely
$$\hat q(900{\,}{\rm MeV})_{\rm case{\,}I}^{(n=10)}=1.26{\,}{\rm GeV}^2/{\rm fm},~~ 
\hat q(900{\,}{\rm MeV})_{\rm case{\,}I}^{(n=50)}=1.39{\,}{\rm GeV}^2/{\rm fm};$$
$$\hat q(900{\,}{\rm MeV})_{\rm case{\,}II}^{(n=10)}=1.78{\,}{\rm GeV}^2/{\rm fm},~~ 
\hat q(900{\,}{\rm MeV})_{\rm case{\,}II}^{(n=50)}=1.98{\,}{\rm GeV}^2/{\rm fm}.$$
We have also addressed the dependence of $\hat q(T)$ on the parameter $\zeta$, which measures the smallness of the energy of thermal gluons with respect to their characteristic spatial momenta. For small values of $\zeta$ under consideration, specifically
$\zeta<\frac{1}{\sqrt{3}}\simeq0.58$, the function $M^2(T)$, Eq.~(\ref{zeta}), describing the imaginary part of the hard thermal loop polarization operator,
is a monotonically increasing function of $\zeta$. Therefore, since $M^2(T)$ enters both the numerator and the denominator of the 
resulting Eq.~(\ref{final}), only a numerical analysis can say whether $\hat q(T)$ increases or decreases with the increase of $\zeta$. We have depicted the results of such an analysis in  
Fig.~\ref{ra5} for case~I and in Fig.~\ref{ra6} for case~II. In case~I, we observe a slow monotonic decrease of $\hat q(T)$ 
with the increase of $\zeta$. In case~II, we observe a small increase of $\hat q(T)$ when $\zeta$ varies from 0.1 to 0.3
and an essentially constant behavior of $\hat q(T)$ with the further increase of $\zeta$.
In both cases, the dependences of $\hat q(T)$ on $\zeta$ are very weak, 
which indicates that the predictions of our model are rather stable.

\section{Concluding remarks}
In this paper, we have evaluated the jet quenching parameter $\hat q$
in SU(3) YM theory.  Our calculation is based on a 
nonperturbative approach, represented by the stochastic vacuum model, which describes 
the scattering of a fast parton off a hard thermal gluon in the leading approximation 
$\propto (g^2\left<(F_{\mu\nu}^a)^2\right>_T)^2$. In our scenario of jet quenching, 
the gluon plasma has two components. The chromo-magnetic condensate
describes the soft component of the gluon plasma, which has been called "epoxy" in the works 
of the Stony Brook group~\cite{epoxy}. The 
hard component of the plasma (with momenta larger than the inverse vacuum correlation length)
is represented in our approach by the hard thermal loop effective theory. In some sense, such a picture of a 
two-component gluon plasma resembles the two-component Landau model of superfluid quantum liquids~\cite{liq}.

In our model, jet quenching originates from Landau damping
of soft gluons by the on-shell hard thermal gluons. The analytic result for $\hat q(T)$ is given by Eq.~(\ref{final}).
Numerically, we have calculated $\hat q(T)$ for two alternative parametrizations of 
the chromo-magnetic condensate and the vacuum correlation length, thereby testing the stability of the result 
against possible uncertainties of these quantities. The numerical results are plotted
in Fig.~\ref{ra4} along with the corresponding fitting curves. Our values of $\hat q$ are somewhat larger than the values obtained 
in pQCD, $\hat q_{\rm pQCD}=1.1\div1.4{\,}{\rm GeV}^2/{\rm fm}$~\cite{baier} and closer to those of other recent 
nonperturbative calculations~\cite{zz, rec}, where $\hat q=1.0\div1.9{\,}{\rm GeV}^2/{\rm fm}$. However, due to the 
difference of our model from that of Ref.~\cite{rec}, our result Eq.~(\ref{final}) differs parametrically from the result of 
that paper. 

Note finally that
our values of the jet quenching parameter are close 
also to the effective values in ${\cal N}=4$ SYM~\cite{Liu:2006he} corresponding to QCD.
Indeed, analyzing possible matchings of the two theories, Gubser~\cite{gubser} has argued
that the result of Ref.~\cite{Liu:2006he} would correspond to the values of the jet quenching parameter of a heavy quark in QCD
$\hat q=0.61\div2.6{\,}{\rm GeV}^2/{\rm fm}$. Since the result of Ref.~\cite{Liu:2006he} is not proportional to the 
Casimir operator $C_R$, the values of the jet quenching parameter for a gluon should be the same. 
These values are of the same order of magnitude as those obtained in the present paper.

\acknowledgments
\noindent
We are grateful to D.~D.~Dietrich for a collaboration at the early
stage of the work. The work of D.A. has been supported through the
contract MEIF-CT-2005-024196.

\end{document}